# Exciton Enhanced Giant Correlated Stoke–Anti-Stokes Scattering of Multiorder Phonons in Semiconductor

Jia-Min Lai[1,2,3], Haonan Chang[1,3], Feilong Song[1], Xiaohong Xu[2], Ping-Heng Tan[1,3], and Jun Zhang[1,3]*

[1]State Key Laboratory of Superlattices and Microstructures, Institute of Semiconductors, Chinese Academy of Sciences, Beijing 100083, China
[2]Key Laboratory of Magnetic Molecules and Magnetic Information Materials of Ministry of Education & Research Institute of Materials Science, Shanxi Normal University, Taiyuan 030031, China
[3]Center of Materials Science and Optoelectronics Engineering, University of Chinese Academy of Sciences, Beijing 100049, China
*Corresponding author. Email: zhangjwill@semi.ac.cn

## Abstract

The correlated Stokes–anti-Stokes (SaS) scattering plays a crucial role in quantum information processing, such as heralded light sources, Fock state dynamics, and write-read protocol for quantum memory. However, several reported materials exhibit low degree of SaS correlation and require high-power pulse laser excitation, limiting further applications. Herein, we explore the giant correlated multiorder SaS scattering under low-power continuous laser excitation through red-sideband resonance of exciton in semiconductor ZnTe nanobelts. At low temperatures, we observe an unexpectedly strong anti-Stokes signal for multiorder longitudinal optical phonons, with SaS correlations two or four orders of magnitude larger than reported results. Furthermore, we observed the mitigation of laser heating effect for longitudinal optical phonon in SaS scattering. This finding paves a new pathway to study multiorder quantum correlated photon pairs produced through exciton-resonant Raman scattering.

## Introduction

Raman scattering, as an inelastic scattering process, provides a powerful tool to study the vibration mode (i.e. phonon) in molecules or crystals. Depending on the energy exchange, the Raman scattering exhibits two components: Stokes and anti-Stokes scattering. In Stokes scattering, the incident photon converts into a lower-energy photon and generates a phonon. While in anti-Stokes scattering, the incident photon converts into a higher-energy photon with a phonon annihilation (Figs. 1(a) and (b)). Therefore, in quantum description, the intensity ratio of anti-Stokes and Stokes scattering depends on the scattering photon frequency and the phonon population[1]: $I_{aS}/I_S = \left(\frac{\omega_L+\Omega}{\omega_L-\Omega}\right)^4 \frac{\bar{n}}{\bar{n}+1}$, where $\omega_L$ is the angular frequency of the pump laser, and $\bar{n} = \left[\exp\left(\frac{\hbar\Omega}{k_B T}\right) - 1\right]^{-1}$ is Bose-Einstein distribution of phonon. The $I_{aS}/I_S$ has been used to investigate the local temperatures of nanostructures [2], thermal conductivity [3], and optical resonances [4]. The general spontaneous Raman scattering is typically regarded as an incoherent process, both Stokes and anti-Stokes intensities are linearly dependent on pump power. However, as depicted in Figs. 1(c) and (d), if the phonon generated by the Stokes process is subsequently annihilated in the anti-Stokes process, the anti-Stokes scattering intensity exhibits a squared dependence on the pump power. Though it has not been reported yet, we could expect that the $I_{aS}$ will surpass $I_S$ at high power excitation. It is distinct from the noncorrelated Raman scattering, where the $I_{aS}/I_S$ approach unity when the temperature is high enough to activate a very large phonon population. Klyshko predicted this nonlinear Raman scattering process called correlated Stokes–anti-Stokes (SaS) scattering in 1977 [5], and it has been demonstrated in graphene [6,7], diamond [8-10], Carbyne chain [11], Rubidium vapor [12], water [13] and other liquid hydrocarbons [14-16].

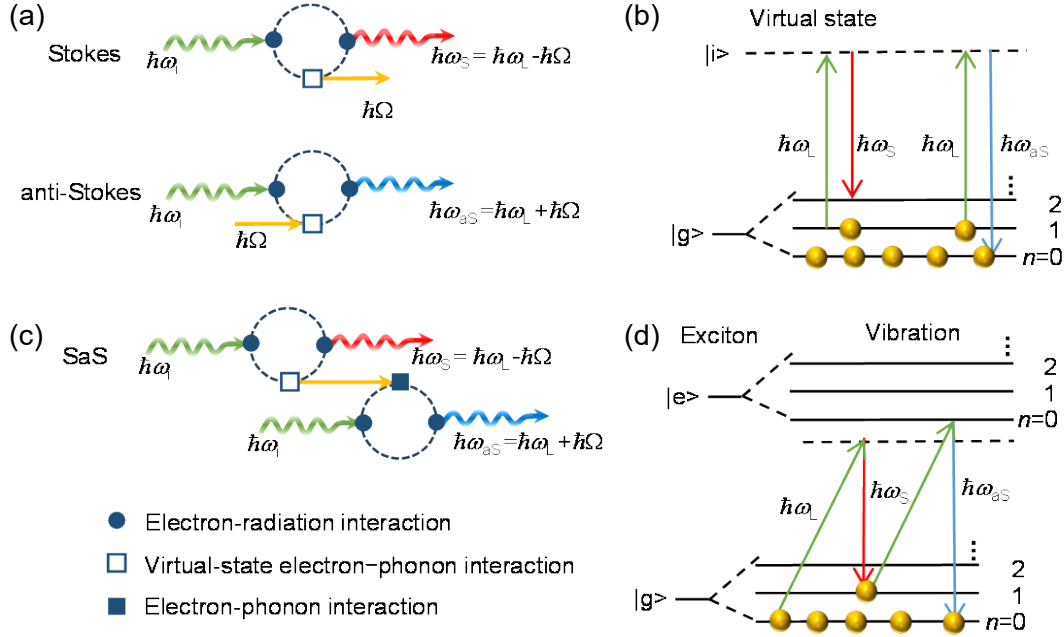

FIG. 1. The principle of the correlated Stoke–anti-Stokes (SaS) scattering. Feynman diagrams for (a) Stokes scattering, anti-Stokes scattering, and (c) SaS scattering. Squiggly lines indicate photons and straight lines indicate phonons. The transition diagram of (b) Stokes scattering, anti-Stoke scattering, and (d) SaS scattering, where |i>, |g>, and |e> represent the virtual state, exciton ground state, and exciton excited state, respectively. |n> denotes the vibration levels.

Similar to the Cooper pairs in the microscopic theory of superconductivity, in SaS scattering, the Stokes and anti-Stokes photons form correlated photon pairs that share the same quantum of vibration [5]. Since the nonclassical correlations of Stokes and anti-Stokes photons were observed in diamond [17], it has been employed in the quantum field. For instance, the measurement of SaS photon correlation constitutes a powerful tool for obtaining heralded single phonons [18-20]. The Stokes and anti-Stokes photons scattered from the "write" and "read" laser pulse, respectively, offer a project for storage and access to vibration quantum state [21-23]. The decay of the SaS correlations maps the dynamics of a single phonon, allowing for measurement of Fock state lifetime [24,25]. However, the degree of correlation in most reported SaS scattering materials is still lower than other sources of photon pairs, such as spontaneous parametric down-conversion [21] or spontaneous four-wave mixing [26,27], which hinders the development of further applications. Therefore, new materials with a large degree of SaS correlation and new physics for enhancement of the SaS correlation are in demand.

There are some common accesses or criteria to obtain SaS photons correlation or coherence by achieving nonlinear power dependence of anti-Stokes signal in Raman scattering, such as pump with high-power pulse laser [6,24,28], surface- or tip- enhancement with plasmon resonance [11,29-31], resonant with a cavity mode [20,32-34], etc. Herein, utilizing the intrinsic exciton resonance, we achieve a correlated SaS scattering of multiorder LO phonon modes with low-power continuous laser excitation in Zinc Telluride (ZnTe) nanobelts, while the other phonon modes remain noncorrelated. At extremely low temperatures, even the anti-Stokes signal of three-order LO phonons can be abnormally observed. The SaS correlation parameter in our experiment is two orders of magnitude larger than that of twisted-bilayer graphene, and four orders of magnitude larger than that of AB-stacked graphene and diamonds [35]. Such giant SaS correlation allowed the anti-Stokes scattering could be observed not only at 4 K but also with an intensity of four times larger than its Stokes counterpart at high power excitation. It is foreseen that these results may open a new door to explore the multiorder correlated SaS photon pairs. Moreover, the dominance of phonon annihilation in the scattering process provides a new way to mitigate the thermal energy at very low temperatures.

**Results and Discussion**

In classical theory, an electromagnetic field $E$ will induce dipole moment ($P$) in atom/molecular and crystal, i.e. $\boldsymbol{P} = \chi^{(1)} \cdot \boldsymbol{E} + \chi^{(2)} : \boldsymbol{EE} + \chi^{(3)} \vdots \boldsymbol{EEE} + \cdots$, where $\chi^{(n)}$ is the electric susceptibility tensor. The normal linear electronic susceptibility tensor $\chi^{(1)}$ characterizes the refraction and absorption process of light by the medium as well as Rayleigh and the spontaneous Raman scattering. The higher-order terms such as $\chi^{(2)}$ and $\chi^{(3)}$ are responses for the nonlinear optical process, including the correlated SaS scattering. The other requirement for the nonlinear optical process is electromagnetic field confinement, which is often realized by the plasma or microcavity resonance in previous experiments [11,29-31]. ZnTe, as a nonlinear crystal for generating and detecting THz radiation [36], owns nonzero $\chi^{(2)}$ and $\chi^{(3)}$ [37]. The laser cooling of single phonon mode (called resolved sideband Raman cooling) has been achieved in ZnTe due to its strong Fröhlich interaction between longitudinal optical (LO) phonon and intrinsic exciton [38]. By analogy, the large electron–phonon coupling also provides the potential opportunity for correlated SaS scattering without other assisted approaches.

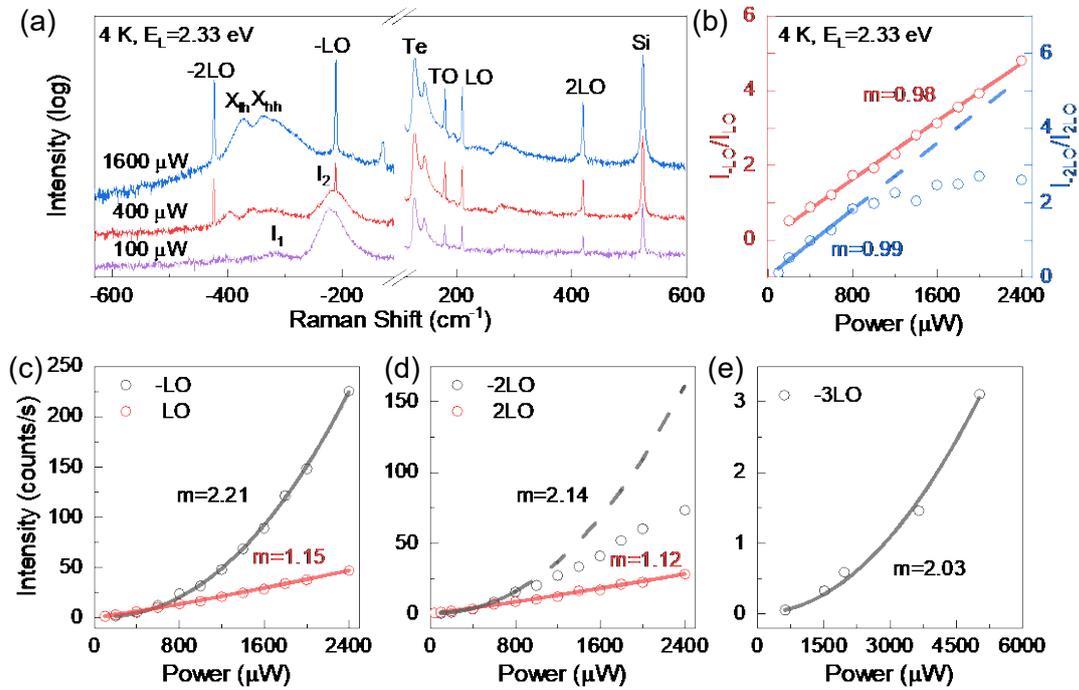

FIG. 2. Power-dependent Raman spectra of ZnTe nanobelts at 4 K. (a) Power-dependent Raman spectra of ZnTe nanobelts excited by 2.33 eV laser. (b) $I_{aS}/I_S$ of LO and 2LO phonons. The line is fitted by $I_{aS}/I_S \propto P^m$. Extracted power dependence of the Raman scattering intensity for (c) LO phonon (d)

2LO phonon, and (e) 3LO phonon. The line is fitted by $I \propto P^m$.

A necessary condition for observing SaS correlation is a small number of phonons excited by temperature fluctuations of the environment [8,35], mitigating the competition with spontaneous anti-Stokes scattering. To effectively utilize the exciton resonance, correlated SaS scattering of ZnTe nanobelt was excited at the red-sideband of the exciton at 4 K (Fig. 2 (a), Figs. S2 and S3). In this case, LO phonons almost fully occupy the ground state ($\bar{n} = 1.10 \times 10^{-33}$). The theoretical thermal equilibrium value of $I_{-LO}/I_{LO}$ for spontaneous Raman scattering is $1.20 \times 10^{-33}$. That is, the spontaneous anti-Stokes scattering hardly occurs at this extremely low temperature. Unexpectedly, the anti-Stokes signal was measured not only for LO phonons but also for 2LO phonons (Fig. 2(a)). In contrast to the resolved-sideband Raman cooling reported at higher temperatures (225 K) [38], the anti-Stokes intensity of LO phonon exhibits quadratic power dependence, while its Stokes counterpart keeps a linear dependence (Fig. 2(c)). Correspondingly, the intensity ratio of anti-Stokes and Stokes scattering of LO phonon ($I_{-LO}/I_{LO}$) shown in Fig. 2(b) exhibits a linear power dependence, which serves as convincing evidence of the correlated SaS scattering. The two-photon process can be excluded due to the low-power excitation we used and the linear dependence on the Stokes intensity of other modes, such as transverse optical (TO) mode and Si Raman signal from the substrate in Fig. S4. Only anti-Stokes signal of LO phonon modes can be observed implies that the phonon mode of SaS correlation scattering can be selected by exciton resonance. This outgoing resonance results from the combined contribution of light hole ($X_{lh}$) and heavy hole ($X_{hh}$) excitons [39], since their energies fall within the scattering energies of LO and 2LO phonons. The PL peaks denoted as $I_1$ and $I_2$ correspond to the excitons bound to the neutral shallow acceptor [39], which become ionized as the power increases and therefore do not contribute to the resonance. It is noted that all exciton PL peaks show a redshift as laser power increases, indicating the local heat-up caused by laser. The anti-Stokes intensity of the 2LO phonon peak deviates from a quadratic power dependence at high power (Fig. 2(d)) due to the deviation from exciton resonance.

Another difference from the previous reports [6-16] is that the anti-Stokes intensity surpasses the Stokes scattering as the pump power increases (Fig. 2(c) and (d)). Under high power excitation, anti-Stokes signal of 3LO mode even appears without any Stokes photons were measured (Fig. 2 (e) and Fig. S3). Similar results were observed in the other two samples in Supplementary Materials (Figs. S5 and S6). These phenomena all indicate the SaS correlation here is larger than that in previous materials [6-16]. Besides, $I_{-LO}/I_{LO}$ that are much larger than the thermal equilibrium value implies that more LO phonons are being annihilated than generated. Ignoring the heat exchange with the thermal bath, the LO phonon can be cooled during the correlated SaS scattering. Two factors should be taken into account for the large $I_{-LO}/I_{LO}$: (1) the exciton resonance effect and (2) the correlated SaS scattering.

To discuss the exciton resonance and the laser heating effect, the temperature-dependent Raman spectra were measured with a low pump power in Fig. 3(a) and Fig. S7. As the temperature increases, the energy of excitons decreases and $X_{lh}$ and $X_{hh}$ become undistinguished. As the exciton energy gradually approaches the energy of LO anti-Stokes scattering, the anti-Stokes intensity of LO exceeds that of Stokes. Subsequently, the exciton energy continues to decrease, deviating from the anti-Stokes resonance, resulting in the reduction in anti-Stokes intensity relative to Stokes intensity. The $I_{-LO}/I_{LO}$ is extracted in Fig. 3(b) with the maximum observed around 70 K, where the anti-Stokes scattering energy of LO phonon accurately matches with the exciton. After passing the anti-Stokes resonance, $I_{-LO}/I_{LO}$ decreases and gradually approaches the thermal equilibrium value (denoted by the blue dashed curve). $I_{-LO}/I_{LO}$ has a Lorentzian profile and can be fitted well with the function:

$$I_{aS}/I_S = k \frac{\bar{n}}{\bar{n}+1} \frac{(E_S-E_X)^2+(\kappa_X/2)^2}{(E_{aS}-E_X)^2+(\kappa_X/2)^2}, \tag{1}$$

where $E_X$ and $\kappa_X$ are the exciton energy and linewidth, $E_S$ and $E_{aS}$ are the energy of the Stokes and anti-Stokes scattering, respectively. $\epsilon = \frac{(E_S-E_X)^2+(\kappa_X/2)^2}{(E_{aS}-E_X)^2+(\kappa_X/2)^2}$ represents the ratio of Stokes and anti-Stokes scattering cross section, which also be defined as a resonance factor. Since the temperature affects the exciton energy and linewidth, the Varshni equation [40] $E_X(T) = E_X(0) - \alpha T^2 / (T + \beta)$ and linear equation $\kappa_X(T) = \kappa_X(0) + cT$ were introduced to describe the energy and linewidth of the exciton, respectively, where $\beta$ is Debye temperature of ZnTe, $\alpha$ and $c$ reflect the temperature coefficient of the exciton shift and linewidth, respectively. By fitting the temperature-dependent $I_{-LO}/I_{LO}$, the Debye temperature of ZnTe is 180 K, consistent with the previous report [41]. Notably, the fitting coefficient $k$ is 13.9, which cannot simply be attributed to the $\omega^4$ dependence of Raman scattering and the frequency dependence of instrumental apparatus. Except for the variation in resonance conditions and phonon number caused by the thermal effect, the quadratic power dependence of anti-Stokes scattering caused by SaS correlation is a significant factor to such large $I_{-LO}/I_{LO}$. The power-dependent results also support this idea. The relation between excitation power and the temperature at the heating area is obtained by comparing the exciton energy in power-dependent with temperature-dependent experiments (see details in Supplementary Materials): $T_X = 4 + 0.19P^{0.65}$. If only the laser heating and resonance effects are considered, the $I_{-LO}/I_{LO}$ at 4 K can be expressed as a function of laser power, as plotted by the blue dots in Fig. 3(c). The substantial deviation from the power-dependent experimental results (red dots) again supports the presence of correlated SaS scattering.

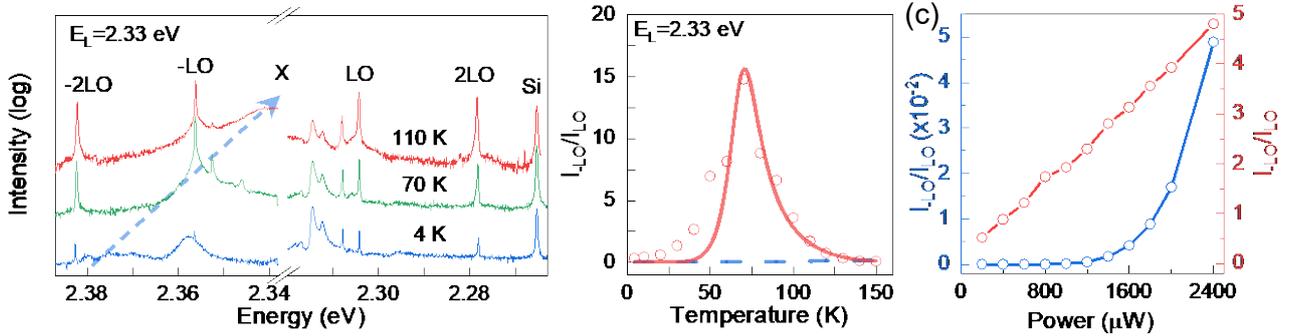

FIG. 3. Temperature-dependent Raman spectra of ZnTe nanobelts. (a) Temperature-dependent Raman spectra excited by 2.33 eV with laser power of 300 μW. (b) The temperature-dependent $I_{aS}/I_S$ of LO phonon excited by 2.33 eV. The blue dash curve denotes the thermal equilibrium value of $I_{-LO}/I_{LO}$, while the red curve is the fitting result with consideration of the resonance factor. (c) The contrast between laser heating-induced theoretical results (blue dots) and correlated SaS scattering dominant experimental results (red dots) of power-dependent $I_{-LO}/I_{LO}$.

Then we evaluate the strength of SaS correlation in ZnTe. The anti-Stokes and Stokes intensity ratio is given by[35]

$$\frac{I_{aS}}{I_S} = \epsilon \frac{\bar{n}}{\bar{n}+1} + C_{SaS}P\left[\frac{1}{\bar{n}+1} - \epsilon \frac{\bar{n}}{(\bar{n}+1)^2}\right]\frac{2r+1}{2r+2} \tag{2}$$

where the first term describes the thermally generated spontaneous Raman scattering, which is independent of laser power. At higher power, the correlated SaS scattering dominants and the $I_{aS}$ is proportional to $P^2$, corresponding to the second term. $C_{SaS}$ gives the effective response of the anti-Stokes line per unit of pump power, which also reflects the strength of SaS correlation. $r = \gamma / \gamma_c$,

$\gamma = \gamma_{aS} = \gamma_S$, where $\gamma_S$, $\gamma_{aS}$, and $\gamma_c$ represent the decay rates (linewidths) of the Stokes, anti-Stokes, and phonon fields, respectively. Here, the excitation photons are generated by a continuous wave laser, thus the scattered photon decay rate $\gamma = \gamma_c$, $r$=1. At 4 K, with the effective phonon population $\bar{n} = 1.10 \times 10^{-33}$, the power-dependent data can be fitted well with Eq. (2) (Fig. 4(a)), yielding the correlation parameter $C_{SaS}$=26.5 $\pm$ 0.2 mW$^{-1}$. This value is two orders of magnitude larger than that of twisted-bilayer graphene ($3.9 \pm 0.3 \times 10^{-2}$ mW$^{-1}$), and four orders of magnitude larger than that of AB-stacked graphene ($7 \pm 8 \times 10^{-4}$ mW$^{-1}$) and diamonds ($1.0 \pm 0.9 \times 10^{-4}$ mW$^{-1}$) reported before [6,10,35]. Such giant correlation enables the observation of correlated SaS scattering for multiorder phonons without pulse laser, and the anti-Stokes process surpasses Stokes process when excited with hundreds of microwatts power. Additionally, the calculated $I_{-LO}/I_{LO}$ with various $\bar{n}$ and $C_{SaS}$=26.5 $\pm$ 0.2 mW$^{-1}$ is presented in Fig. 4(b). As the phonon population $\bar{n}$ increases, the pump power required for correlated SaS scattering increases, which explains the occurrence of correlated SaS process at 4 K. Regrettably, the comparable quadratic power dependence of $I_{-LO}$ was not observed at 70 K (Fig. S8), presumably due to the variation in the correlation coefficient $C_{SaS}$ that necessitates a higher pump power for correlated SaS scattering.

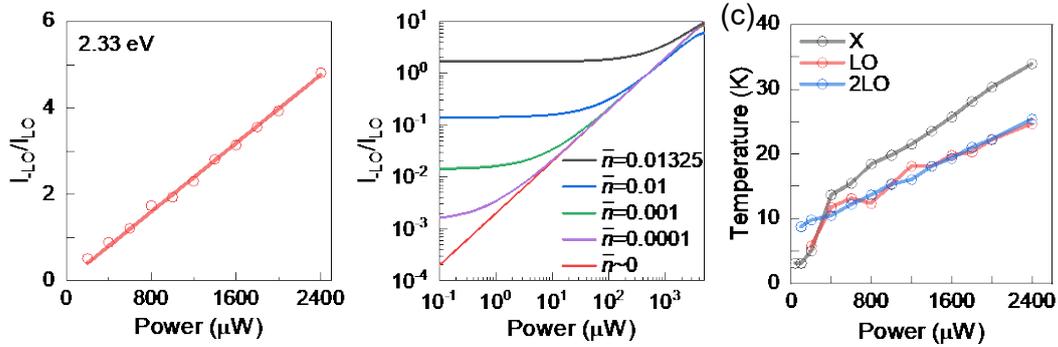

FIG. 4. Theoretical description of the $I_{aS}/I_S$ and evaluation of LO phonon temperature. (a) The power-dependent $I_{-LO}/I_{LO}$ fitted with Eq. (2). (b) The calculated $I_{-LO}/I_{LO}$ as a function of power and phonon population $\bar{n}$. (c) The temperature evolution of the exciton, LO phonon, and 2LO phonon as a function of laser power.

With the laser power increases, $I_{-LO}/I_{LO}$ far exceeds the thermal equilibrium value, indicating that more LO phonons are annihilated than generated. A. Jorio et al. have briefly discussed the existence of the cooling term in SaS scattering[35], but up to now there have been no experimental reports of SaS-based cooling so far. The second term in Eq. (2) cools down the phonon when the anti-Stokes scattering surpasses the Stokes scattering. To evaluate the cooling effect, the phonon energy as a function of temperature was extracted in Figs. S9 and S10. By mapping the peak position of the power-dependent spectra to the temperature, the temperature of LO phonon $T_{LO}$ as a function of laser power is depicted in Fig. 4(c). The sample temperature can be represented by the exciton temperature $T_X$. The temperature of LO phonon obtained from the LO and 2LO peaks exhibits consistency and is lower than the exciton temperature $T_X$ at the same laser power. It shows that although the laser has a heating effect on the whole sample, this heating effect on the LO phonons is relatively suppressed. In other words, similar to the sideband Raman cooling [38,42], the correlated SaS scattering can also cool (or manipulate) an individual vibration mode. The normalized temperature difference $\Delta T/P$ is defined as $\Delta T/P = \frac{T_{LO}-T_X}{P} = -\xi\alpha\eta$, where $\alpha$ is the absorption coefficient (~$10^4$ cm$^{-1}$ for ZnTe[39]), $\eta$ denotes the cooling efficiency of LO phonons. $\xi$ represents the thermal resistance coefficient of the phonon and thermal bath, which is associatied with the exciton–phonon coupling and phonon–phonon anharmonic effect[43]. It can be inferred that in ZnTe, a large thermal resistance exists between the LO phonon and thermal bath, and accordingly, the correlated SaS scattering can mitigate the laser heating of a single phonon mode. Moreover, it can be hypothesized that if the thermal resistance between the phonon and thermal bath is minimal, the entire sample might be cooled as long as the laser pumps for a sufficiently extended duration.

**Conclusion**

In summary, we demonstrate a correlated SaS scattering in ZnTe at 4 K with a giant correlated parameter. Different from previous reports, the anti-Stokes signal of multiphonon modes were measured in this work. As the laser power increases, the intensity of anti-Stokes component takes on a more prominent role and overshadows the Stokes component. As a result, the temperature of LO phonon remains lower than that of the local temperature at the laser spot, indicating mitigation of the laser heating effect on LO phonon. Our work contributes to a deeper comprehension of correlated SaS scattering at ultra-low temperatures and proposes a new mechanism for phonon mode manipulation by laser resonance. Looking forward, these results hold promise for extension to the fields of cold atoms, molecules, and cavity optomechanics to develop the manipulation and storage of quantum states and avoid decoherence of the vibration state induced by the laser heating effect.

**Competing interests**

The authors declare no competing financial interest.

**Acknowledgments**

J. Z. acknowledges the CAS Project for Young Scientists in Basic Research (YSBR-120), National Natural Science Foundation of China (12074371), Research Equipment Development Project of Chinese Academy of Sciences (YJKYYQ20210001). J.-M. L. acknowledges the National Natural Science Foundation of China (12404091).

# Supplementary Materials for

# Exciton Enhanced Giant Correlated Stoke–Anti-Stokes Scattering of Multiorder Phonons in Semiconductor

Jia-Min Lai[1,2,3], Haonan Chang[1,3], Feilong Song[1], Xiaohong Xu[2], Ping-Heng Tan[1,3], and Jun Zhang[1,3]*

[1]State Key Laboratory of Superlattices and Microstructures, Institute of Semiconductors, Chinese Academy of Sciences, Beijing 100083, China

[2]Key Laboratory of Magnetic Molecules and Magnetic Information Materials of Ministry of Education & Research Institute of Materials Science, Shanxi Normal University, Taiyuan 030031, China

[3]Center of Materials Science and Optoelectronics Engineering, University of Chinese Academy of Sciences, Beijing 100049, China

*Corresponding author. Email: zhangjwill@semi.ac.cn

## Methods

The ZnTe nanobelts were synthesized in a home-built, vapor-transport system for chemical-vapor deposition and then dispersed into single nanobelt on 90 nm $SiO_2$/Si substrates. Raman measurements were collected in backscattering geometry with a confocal triple-grating spectrometer (Horiba-JY T64000) equipped with a liquid-nitrogen-cooled charge-coupled detector (CCD). The corresponding spectral resolution is 0.64 $cm^{-1}$ per CCD pixel. A single frequency continuous wave laser with wavelength of 532 nm is focused by a high numerical aperture (0.82) objective lens. The low temperature is provided by a closed-cycle cryostat (attoDRY 800).

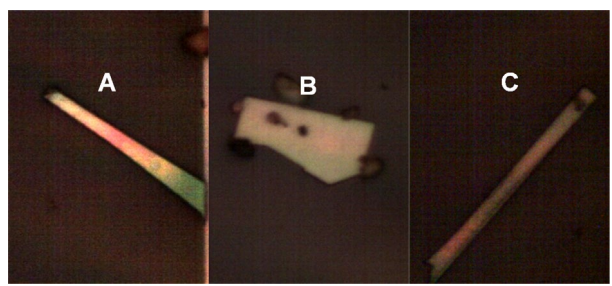

FIG. S1. Optical microscope image of ZnTe nanobelts marked as Sample A, B, C. The experimental data in the main text are from Sample A.

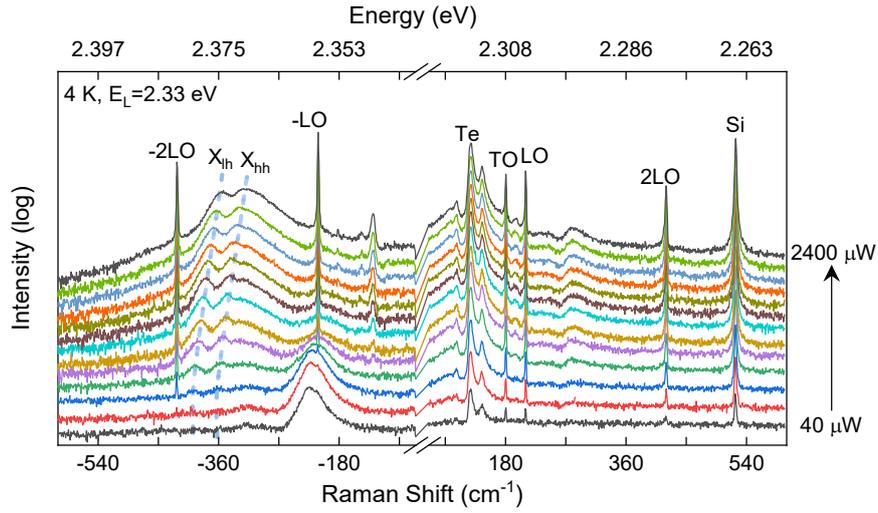

FIG. S2. Laser power-dependent Raman spectra of ZnTe nanobelt (Sample A) at 4 K.

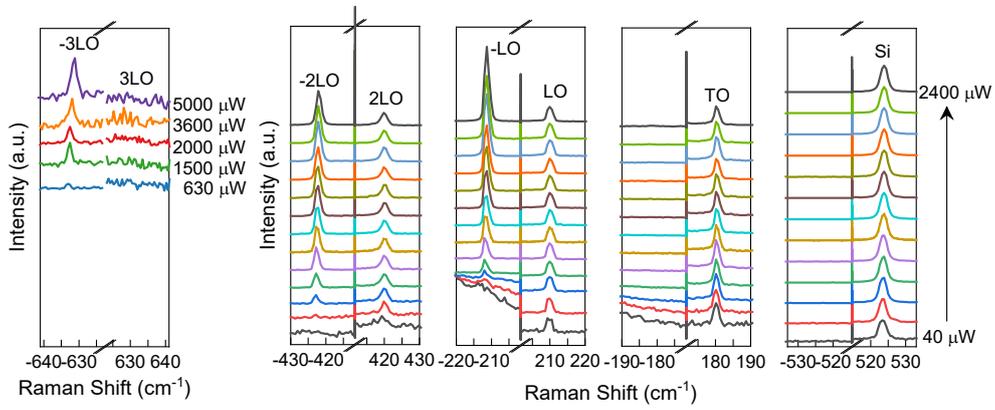

FIG. S3. Anti-Stokes/Stokes Raman spectra of ZnTe nanobelt (Sample A) at 4 K.

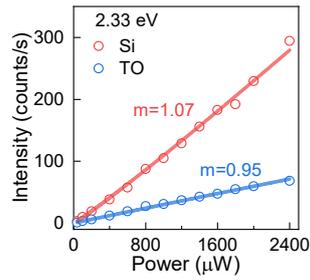

FIG. S4. Laser power dependence of Raman scattering intensity for Si signal and TO phonon of Sample A in Figs. S2 and S3 at 4 K.

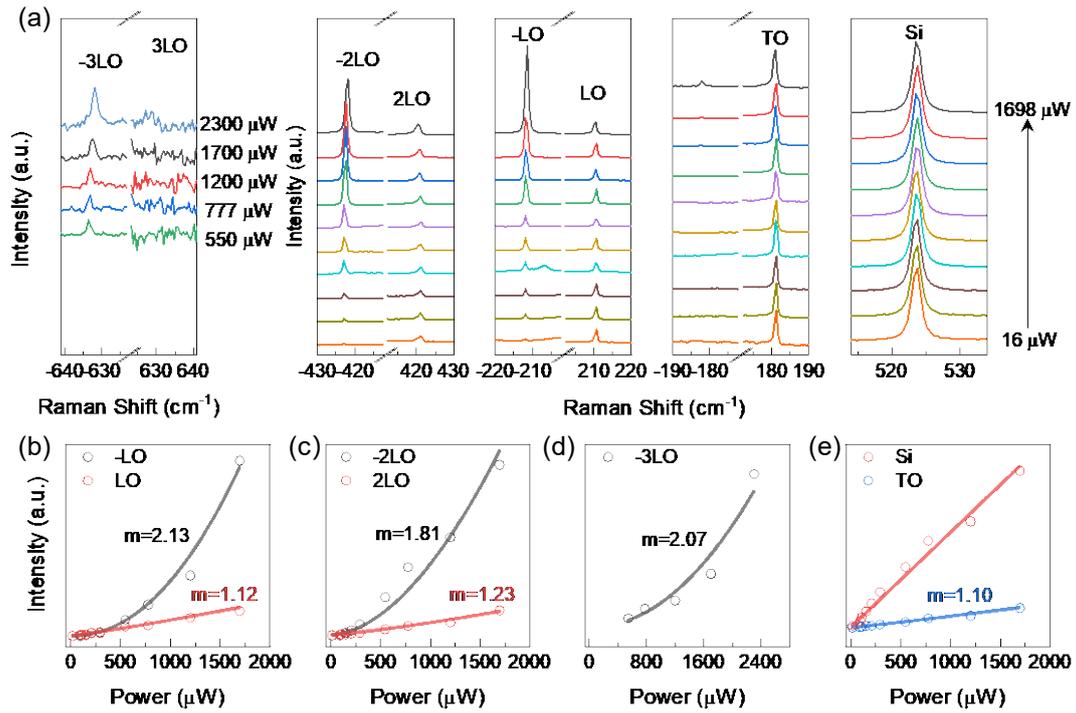

FIG. S5. Anti-Stokes/Stokes Raman spectra of ZnTe nanobelt (Sample B) at 4 K.

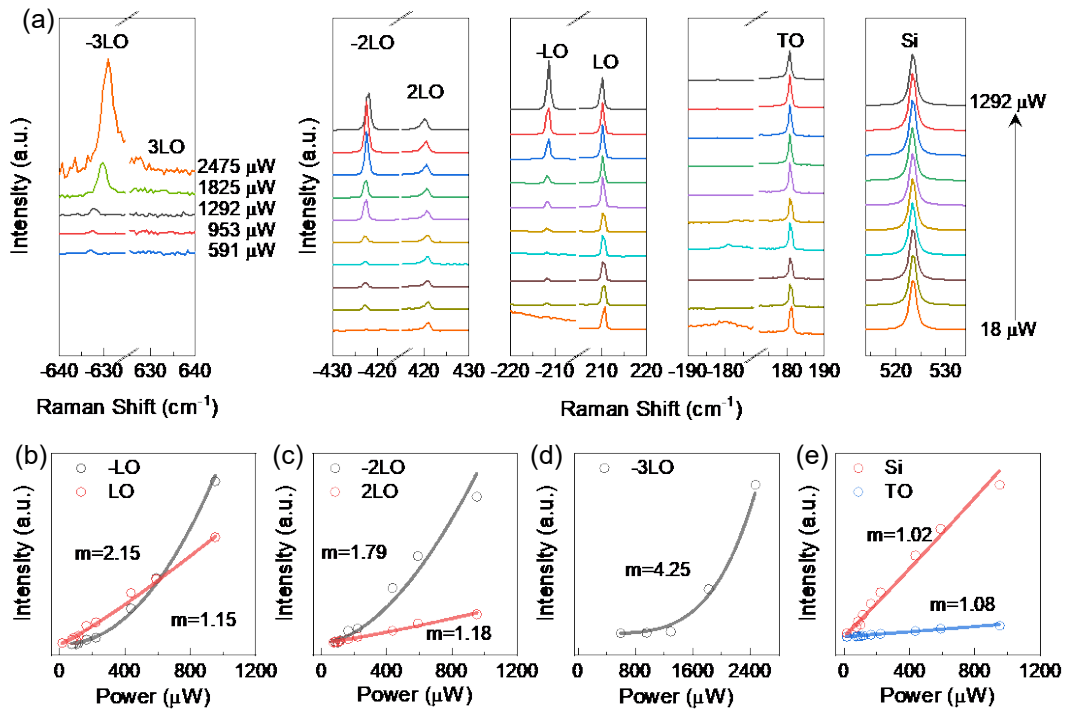

FIG. S6. Anti-Stokes/Stokes Raman spectra of ZnTe nanobelt (Sample C) at 4 K.

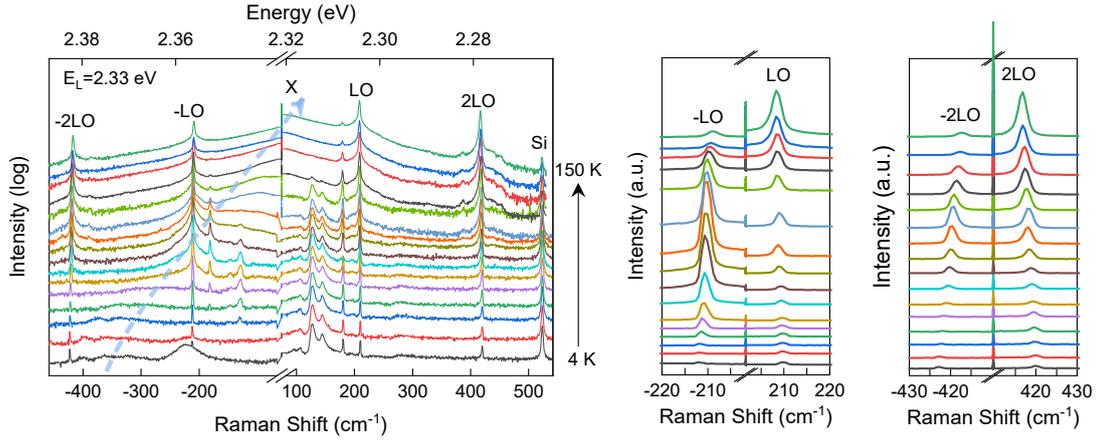

FIG. S7. Temperature-dependent Raman spectra of ZnTe nanobelt (Sample A).

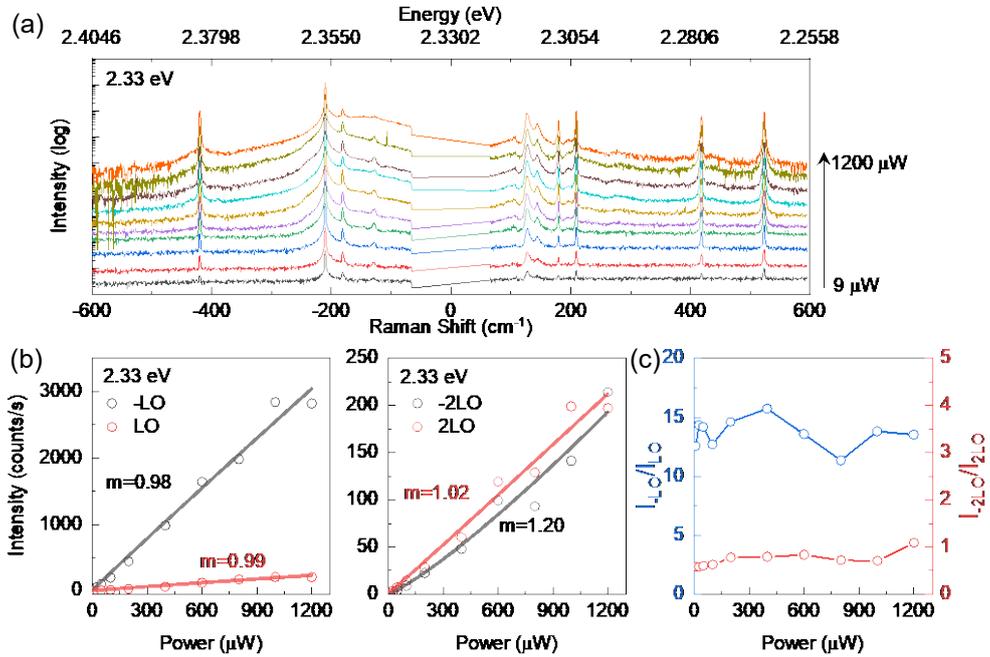

FIG. S8. Laser power-dependent Raman spectra of ZnTe nanobelts at 70 K. (a) Power-dependent Raman spectra of ZnTe nanobelts excited by 2.33 eV (532 nm) laser. (b) Extracted power dependence of the Raman scattering intensity for LO and 2LO phonon. The line is fitted by $I \propto P^m$. (e) $I_{aS}/I_S$ of LO and 2LO phonons.

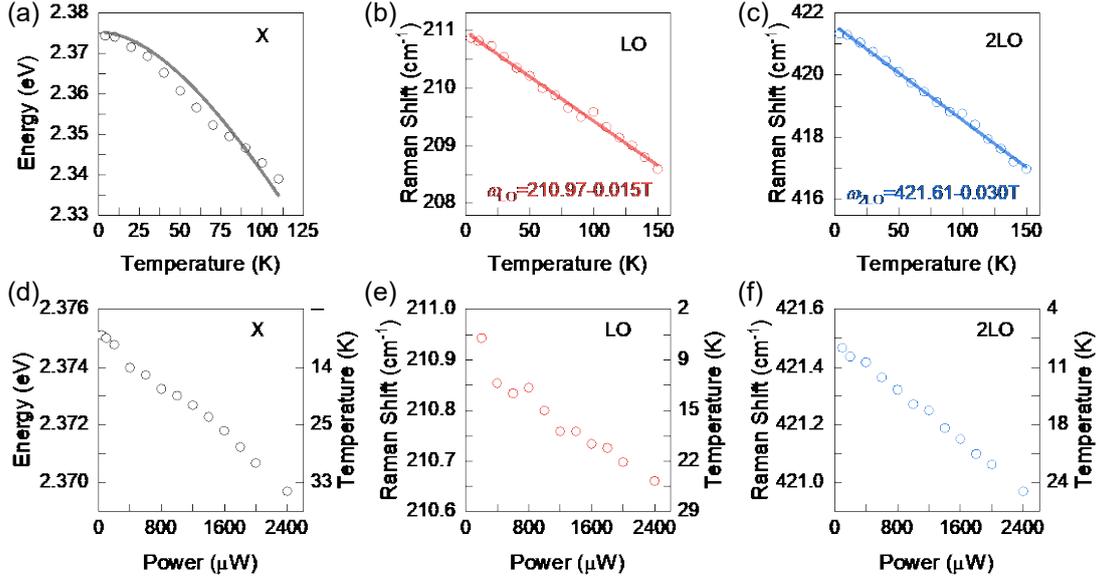

FIG. S9. Peak position of (a) exciton, (b) LO phonon, and (c) 2LO phonon in temperature-dependent spectra. The effective temperatures of (d) exciton, (e) LO phonon, and (f) 2LO phonon in power-dependent spectra.

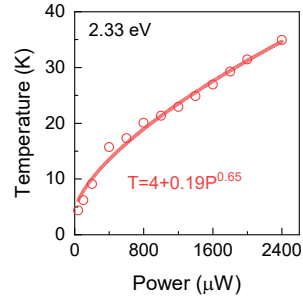

FIG. S10. Effective sample temperature induced by laser heating effect. The red line is fitted with $T_X = 4 + 0.19P^{0.65}$.

1. **Laser heating effect**

As seen in Fig. S1, the red shift of exciton energy is attributed to laser heating effect. When the laser beam is focused on the surface of ZnTe, the energy in the nonradiative transition process is transformed into heat energy. According to the work of Yang et al.[1], under an excitation power of $P$, the power density of heat $P_h$ can be expressed as

$$P_h = K(1-\eta)FP, \quad (S1)$$

Here, $\eta$ is the total luminescence quantum yield when the sample is being irradiated with an excitation power of $P$. $F$ is the fraction of incident light absorbed by ZnTe nanobelts and $K$ is a proportionality constant. When laser heating and ambient heat exchange are in balance, there is a relation between the power density of heat ($P_h$) and the temperature at the heating area ($T$):

$$P_h \approx \kappa(T - T_0), \quad (S2)$$

here $\kappa$ is the coefficient, which is directly proportional to the thermal conductivity, and $T_0$ is the sample holder temperature. Thus, the temperature of ZnTe under laser heating effect is expressed as

$$T \approx T_0 + \mu P, \quad (S3)$$

where $\mu = K(1-\eta)F/\kappa$. Under the same laser power, a larger $\mu$ leads to a stronger local heating effect. In Fig. S4, we fit the data with a sublinear function ($T_X = 4+0.19P^{0.65}$) very well. The main reason is that the thermal conductivity increases with temperature.

[1] Y. Yang, H. Yan, Z. Fu, B. Yang, L. Xia, Y. Xu, J. Zuo, and F. Li, J. Phy. Chem. B **110**, 846 (2005).